\documentclass[11pt]{article}

\usepackage{amsmath,amssymb}
\usepackage{graphicx}
\usepackage{geometry}
\usepackage[inline]{enumitem}

\usepackage{hyperref}
\usepackage{natbib}

\providecommand{\keywords}[1]
{
  \small	
  \textbf{\textit{Keywords---}} #1
}

\title{LLAssist: Simple Tools for Automating Literature Review Using Large Language Models}

\author{Christoforus Yoga Haryanto\\
School of Science, RMIT University, Melbourne, VIC 3000, Australia}

\date{}

\begin{document}

\maketitle

\begin{abstract}
This paper introduces LLAssist, an open-source tool designed to streamline literature reviews in academic research. In an era of exponential growth in scientific publications, researchers face mounting challenges in efficiently processing vast volumes of literature. LLAssist addresses this issue by leveraging Large Language Models (LLMs) and Natural Language Processing (NLP) techniques to automate key aspects of the review process. Specifically, it extracts important information from research articles and evaluates their relevance to user-defined research questions. The goal of LLAssist is to significantly reduce the time and effort required for comprehensive literature reviews, allowing researchers to focus more on analyzing and synthesizing information rather than on initial screening tasks. By automating parts of the literature review workflow, LLAssist aims to help researchers manage the growing volume of academic publications more efficiently.
\end{abstract}


\section{INTRODUCTION}
\label{sec:introduction}
The landscape of academic research is undergoing a dramatic transformation, driven by an unprecedented surge in scientific publications. Identifying relevant work, once a manageable task, has evolved into a time-consuming and often overwhelming process \citep{Khan2003Five,Davis2014Viewing}. The exponential growth of scientific publications \citep{SilvaJúnior2021roadmap,Ioannidis2021rapid}, the need to screen hundreds or thousands of articles, and pressure for rapid-evidence gathering present significant challenges for researchers, potentially compromising research quality \citep{McDermott2024quality,Glasziou2020Waste,Whiting2016ROBIS:,Page2021PRISMA,Wallace2010Semi-automated}.

In response to these mounting challenges, the research community has begun exploring automated solutions to assist in the systematic literature review process \citep{SilvaJúnior2021roadmap,Wallace2010Semi-automated,Tsafnat2018Automated,Bannach-Brown2019Machine}. Recently, there has been a growing interest in evaluating the potential of Large Language Models (LLMs) for this purpose \citep{Agarwal2024LitLLM:,Joos2024Cutting,Susnjak2023PRISMA-DFLLM:}. These advanced AI systems, with their ability to understand and generate human-like text, offer promising avenues for automating various aspects of the literature review process. Yet, various consumer LLM interfaces such as ChatGPT encountered significant challenges in ensuring reliability and consistency of outputs, adhering to rigorous systematic review methodologies, and addressing ethical concerns related to academic integrity \citep{CongLem2024Systematic}.

Without transparent, open-source tools for AI-assisted literature review, researchers face potential limitations in research transparency and replicability that potentially perpetuate biases. These challenges include overlooking critical publications due to time constraints, decreased accuracy from cognitive overload during manual screening, and compromised reproducibility from reliance on closed, proprietary AI systems. Current approaches, while promising, have limitations: \citet{Agarwal2024LitLLM:} focused on generating related work sections but not initial screening; \citet{Joos2024Cutting} explored computationally intensive multi-agent approaches; and \citet{Susnjak2023PRISMA-DFLLM:} proposed resource-intensive domain-specific fine-tuned LLMs. LLAssist addresses these issues by providing a simple, transparent alternative that researchers can freely modify and adapt.

A notable contribution to this emerging field comes from \citet{Joos2024Cutting}, who recently published an extensive evaluation of using LLMs in enhancing the screening process, with results indicating promising potential for reducing human workload. Inspired by these findings, we present LLAssist, a prototype automation tool based on LLM technology. We aim LLAssist to be a building block for an automated knowledge base and research platform, and the open-source nature enables researchers to expand from the idea. This design choice allows adaptation to different research domains with transparent and interpretable results. The simplicity of its design is intentional to allow researchers to understand, modify, and build upon its core functionalities.

Our key contributions include:
\begin{enumerate*}
    \item introducing LLAssist, an open-source tool leveraging LLMs to automate key aspects of the literature review process;
    \item demonstrating a novel approach to relevance estimation using LLMs;
    \item providing insights into different LLM backends' performance for literature review tasks; and
    \item promoting transparency and reproducibility in AI-assisted literature reviews through open-source development.
\end{enumerate*}

\section{METHODOLOGY}
\label{sec:methodology}
Our methodology addresses the primary research question: How can automation of screening using large language models improve the efficiency and effectiveness of systematic literature reviews in the face of exponentially growing scientific publications?

It consists of two main parts:
\begin{enumerate*}
  \item the design and implementation of LLAssist, and
  \item the experimental evaluation of its performance.
\end{enumerate*}

\subsection{LLAssist: An LLM-based Literature Screening Tool}
\subsubsection{Data Input}
The program accepts two primary inputs: a CSV file containing tabular metadata and abstracts of research articles, and a text file listing the research questions of interest. While the current implementation parses the file programmatically, the progress of LLM may allow it to process tabular data reliably despite the current challenges \citep{Dong2024Large}.

\subsubsection{Article Processing}
For each article in the input CSV, LLAssist performs the following steps:

\paragraph{Key Semantics Extraction}
The NLPService extracts topics, entities, and keywords from the article's title and abstract. The implementation sends an extraction prompt with the title and abstract to the LLM.

\paragraph{Relevance Estimation}
For each research question provided, LLAssist estimates the article's relevance using the following criteria:
\begin{enumerate}
    \item \textbf{Binary Relevance Decision and Score:} A binary TRUE/FALSE and a numerical score (0-1) to indicate how closely the article aligns with the research question,
    \item \textbf{Binary Contribution Decision and Score:} A binary TRUE/FALSE and a numerical score (0-1) to assess the article’s potential contribution in answering the research question,
    \item \textbf{Relevance Reasoning:} A brief explanation of why the article is considered relevant.
    \item \textbf{Contribution Reasoning:} A justification for the estimated contribution of the article. 
\end{enumerate}
The current implementation sends the title, abstract, and key semantics previously extracted. An article is considered relevant or contributing if its score exceeds 0.7. This threshold can be adjusted based on the researcher's needs.

\paragraph{Must-Read Determination}
Based on the relevance and contribution scores across all research questions, LLAssist determines whether an article is a "must-read." Currently, this is implemented by using the logical OR operation on all the RQ relevance and contribution thresholds. This binary classification helps researchers prioritize their reading list and conclude the flow.
\paragraph{Output Generation}
LLAssist provides two types of output:
\begin{enumerate*}
    \item a JSON file containing detailed information for each processed article, including extracted semantics, relevance scores, and reasoning, and
    \item a CSV file presenting the same information in a tabular format, suitable for further analysis or import into other tools.
\end{enumerate*}
Necessitating other analysis and tools is deliberate to ensure that the human-in-the-loop principle is adhered to by maintaining the visibility of the process \citep{Buçinca2021To,Vasconcelos2023Explanations,Bansal2021Does}.

\subsection{Experimental Evaluation}
\subsubsection{Data Collection}

To evaluate the effectiveness of LLAssist in streamlining the literature review process for LLM applications in cybersecurity, we conducted two separate experiments using manual sampling of publications from datasets from different sources:
\begin{enumerate}
    \item \textbf{IEEE Xplore Database:} using the search query \textit{"llm AND cyber AND security"} (term a), effectively focusing on recent publications for initial system testing.
    \item \textbf{Scopus Database:} using the search queries \textit{“llm AND cyber AND security”} (term b), \textit{“llm OR ( generative AND artificial AND intelligence ) AND cyber AND security”} (term c), and a broad \textit{“artificial AND intelligence AND cyber AND security”} – limited to Conference paper – (term d) to obtain a diverse dataset in a consecutively larger sample size.
\end{enumerate}

The experiments were designed to assess LLAssist's performance across different academic databases and to ensure a comprehensive evaluation of its capabilities in the fields already known by the authors. There will be two parts: a small dataset test and a large dataset test. The small dataset test is to help manually verify the functionality of LLAssist and the large dataset test is to confirm its utility in enhancing the screening in structured literature review.

From the IEEE Xplore dataset, we exported 17 relevant articles (term a). The Scopus dataset provided an additional set of 37 (term b), 115 (term c), and 2,576 (term d) articles, expanding our corpus. Each dataset included metadata such as titles, abstracts, authors, publication venues, and keywords. The data was exported using the CSV export function to ensure consistency and compatibility with LLAssist's input requirements.

\subsubsection{Research Questions}
We formulated four key research questions to guide our automated analysis:
\begin{itemize}
    \item RQ1: How are Large Language Models (LLMs) being utilized to enhance threat detection and analysis in cybersecurity applications?
    \item RQ2: What are the potential risks and vulnerabilities introduced by integrating LLMs into cybersecurity tools and frameworks?
    \item RQ3: How effective are LLM-based approaches in generating and detecting adversarial examples or evasive malware compared to traditional methods?
    \item RQ4: What ethical considerations and privacy concerns arise from using LLMs to analyze and process sensitive cybersecurity data?
\end{itemize}

\subsubsection{Automated Analysis}
We processed each paper through LLAssist, which performed the following tasks:
\begin{enumerate*}
    \item Extract key semantics (topics, entities, and keywords) from the title and abstract.
    \item Evaluate the relevance of each paper to our research questions.
    \item Provide relevance and contribution scores (0-1 scale) for each research question.
    \item Generate reasoning for relevance and contribution assessments.
\end{enumerate*}
Section \ref{sec:technical implementation} details the technical implementation of the automated analysis done by the program.

\subsubsection{Evaluation Metrics}
We assessed the performance of LLAssist based on i) consistency of evaluations across papers, ii) accuracy in matching papers to relevant research questions, and iii) ability to provide meaningful insights and reasoning. Additionally, we are using several different LLM backends: Llama 3:8B, Gemma 2:9B, GPT-3.5-turbo-0125, and GPT-4o-2024-05-13 to allow data comparison.
\subsubsection{Preliminary Nature of Evaluation}
It is important to note that the assessment of accuracy in matching papers to relevant research questions and the ability to provide meaningful insights and reasoning was conducted in an uncontrolled environment. Knowing that LLM results can be helpful yet inaccurate, we expect the researchers to use LLAssist as a lightweight filtering enhancement tool while following existing methodologies such as PRISMA \citep{Page2021PRISMA,Susnjak2023PRISMA-DFLLM:}.

\section{TECHNICAL IMPLEMENTATION}
\label{sec:technical implementation}
Chain-of-Thought (CoT) prompting is a technique that enhances the reasoning capabilities of large language models by encouraging step-by-step thinking \cite{Wei2022Chain}. In LLAssist, CoT is simulated through two main steps:
\begin{enumerate*}
    \item Extract Key Semantics: The model helps to generate intermediate reasoning steps, expanding on key concepts in the prompt.
    \item Filtering: From the expanded reasoning chains, the most consistent and logical path will be selected as part of the self-consistency in the CoT \cite{Wang2022Self}.
\end{enumerate*}
This approach allows LLAssist to break down complex problems into manageable steps, mimicking human-like reasoning.

LLAssist's approach to literature review automation shares similarities with recent works but also introduces distinct features. Like the LitLLM system proposed by \citet{Agarwal2024LitLLM:}, LLAssist employs a modular pipeline that leverages LLMs for key tasks. However, while LitLLM focuses on generating related work sections, LLAssist prioritizes efficient screening and relevance assessment. Compared to the approach described by \citet{Joos2024Cutting}, which focuses on employing multiple LLM agents for consensus-based filtering, LLAssist currently utilizes a single LLM for both semantics extraction and relevance assessment. While this may sacrifice some of the robustness gained from multi-agent consensus, it simplifies the implementation and potentially reduces processing time.

\begin{figure}[!ht]
    \centering
    \includegraphics[width=0.7\linewidth]{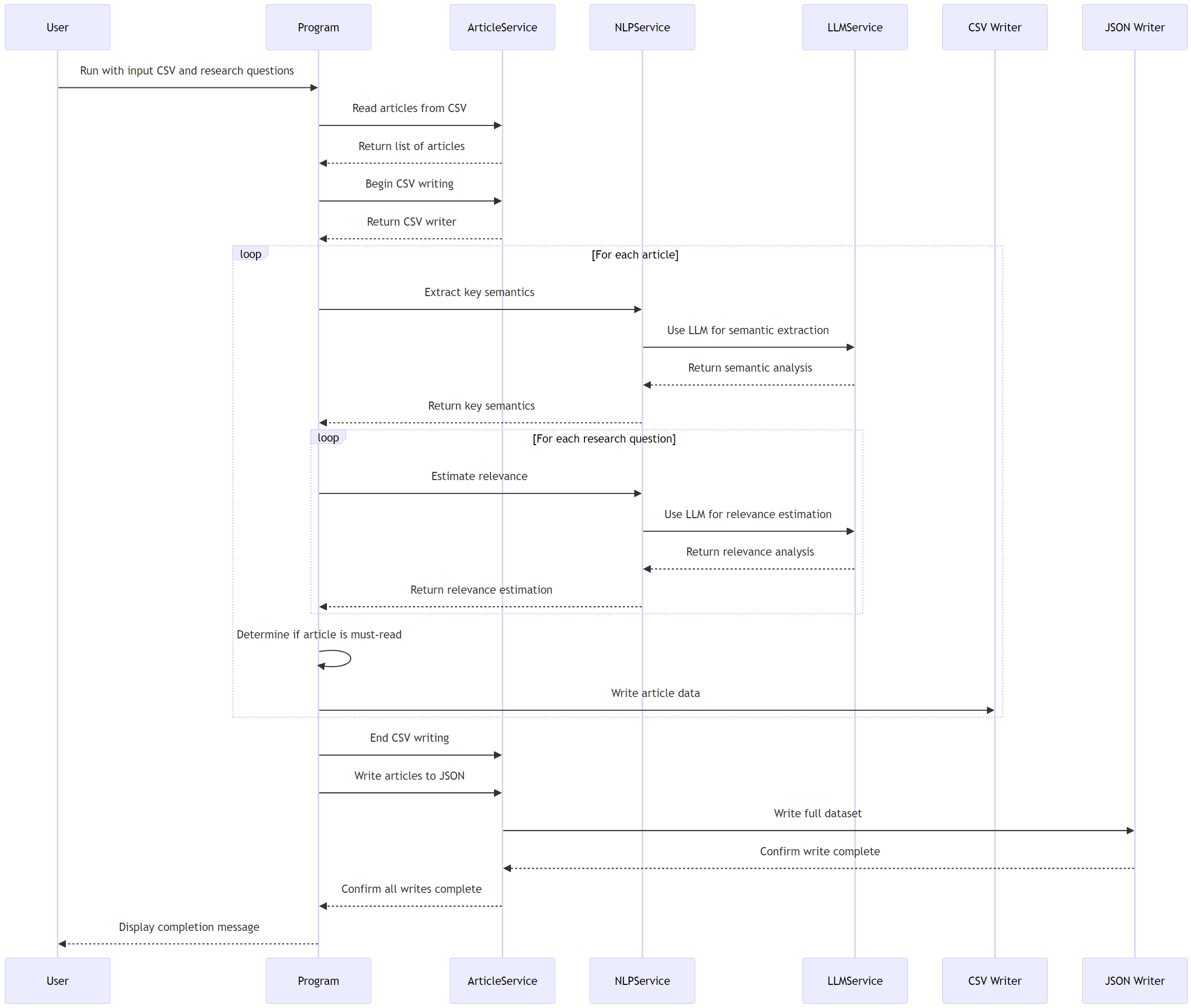}
    \caption{Sequence diagram of LLAssist console application}
    \label{fig:figure1}
\end{figure}

The program is designed to work with various LLM providers, including local models (e.g., Ollama Llama 3, Ollama Gemma 2) and cloud-based models (e.g., OpenAI's GPT-3.5 and GPT-4). This flexibility allows researchers to choose models based on their specific requirements, such as processing speed, accuracy, or data privacy concerns. LLAssist is developed in C\#, a static and strongly typed language \citep{Nanz2015Comparative}, suitable for further integration into a bigger enterprise system. The sequence diagram can be seen in Figure \ref{fig:figure1}.

\section{EXPERIMENT RESULTS}
\label{sec:experiment results}
There are two parts to this experiments:
\begin{enumerate*}
    \item A small dataset test using search term a: IES, term b: SS, term c: SM.
    \item A large dataset test using search term d: SL.
\end{enumerate*}
The small dataset test uses 4 LLMs: Gemma 2, GPT-3.5, GPT-4o, and Llama 3 and the large dataset test only use 2 LLMs: Gemma 2 and Llama 3, both provisioned locally using Ollama on RTX 3090. Time and cost data are measured indirectly from file time and API usage.

\subsection{Small Dataset Test}
The small dataset verified the functionality of the system with the following key results:
\subsubsection{Key Semantics Extraction}
LLAssist successfully identified relevant topics, entities, and keywords for each paper, aligning well with the author-provided keywords and terms for all LLM backends. As this data is currently used as the input to the LLM prompt, there is no controlled measurement done in this experiment. Result data can be reviewed in the CSV. Also, note that the metadata for the classifier is generated by the same LLM.
\subsubsection{Binary Relevance Decision and Score Distribution}
\begin{figure}[!ht]
    \centering
    \includegraphics[width=1\linewidth]{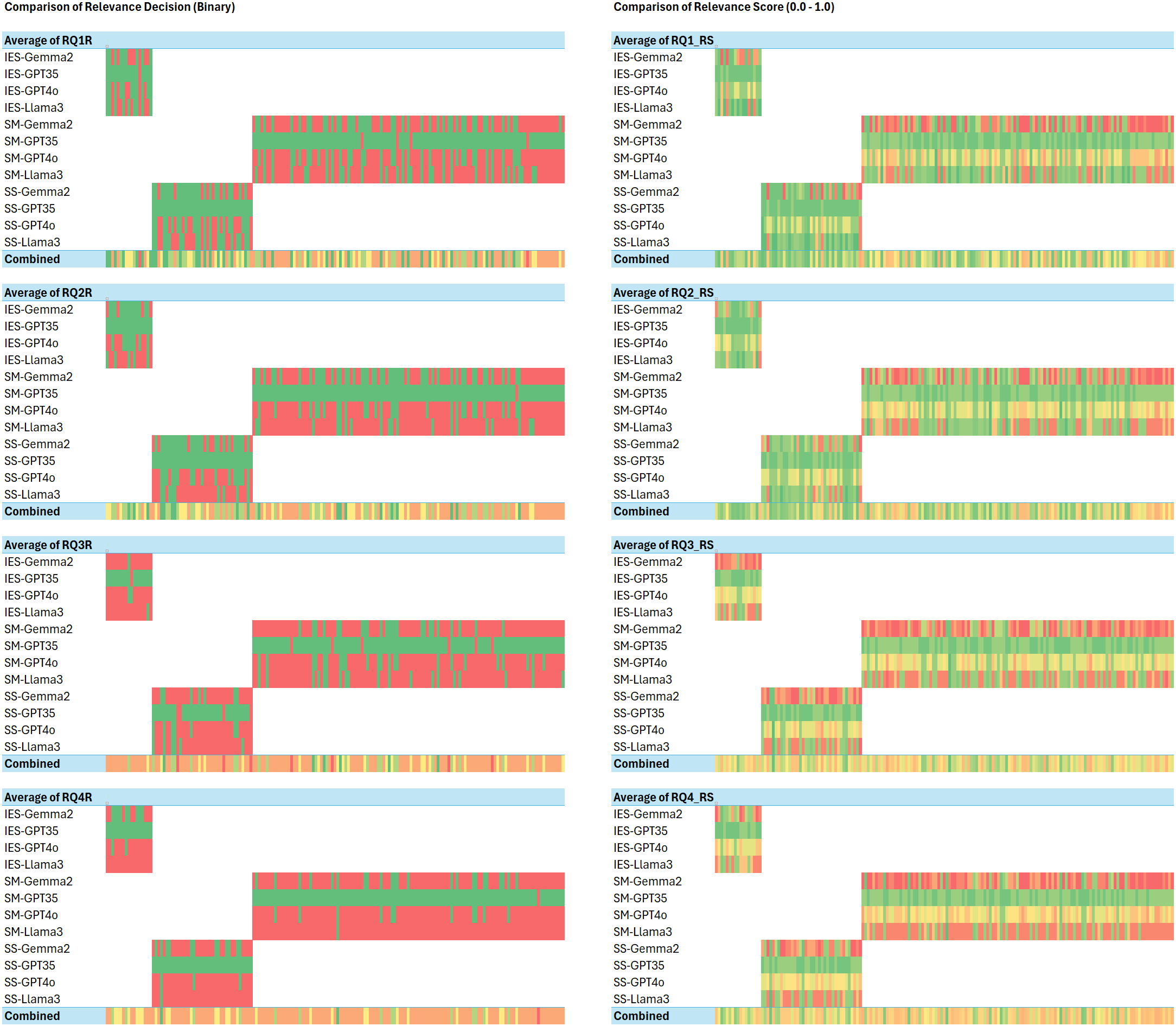}
    \caption{Binary Relevance and Relevance Score Distribution (small datasets)}
    \label{fig:figure2}
\end{figure}

The binary relevance decision and relevance score distribution obtained from the experiment are shown in Figure \ref{fig:figure2}. Further, the summary of binary relevance and binary contribution decisions is shown in Table 1.
\begin{enumerate}
    \item \textbf{Gemma 2:9B} shows a reasonable distribution of binary relevance classification and relevance score. It tends to give a strong binary decision and classification score compared to all other LLMs with variation across research questions, indicating sensitivity to different topics.
    \item \textbf{GPT-3.5-turbo-0125} consistently shows high relevance scores and classifications across all research questions and datasets. It demonstrates the least discrimination among the models, potentially overestimating relevance. This behavior suggests a very inclusive interpretation of research questions, which could lead to a high false positive rate in article selection.
    \item \textbf{GPT-4o-2024-05-13} demonstrates a more balanced distribution between relevant and non-relevant articles. It appears more selective than GPT-3.5 and slightly more selective than Gemma 2, yet more permissive than Llama3 in binary classifications. GPT-4o also tends to return a middle-ground score: while it may imply a sophisticated evaluation, it may also indicate avoiding judgments.
    \item \textbf{Llama 3:8B} exhibits a significant discrepancy between its relevance scores and binary classifications. In binary classification, it's the most conservative and frequently marking articles as not relevant. However, its relevance scores show more diversity, with a range of values that don't always align with its binary decisions. This inconsistency suggests potential issues in threshold setting or score-to-classification conversion for this model, hence we decided to not use the binary relevance decisions from Llama 3 as the basis for the literature screening performance analysis.
\end{enumerate}

\subsubsection{Must-read vs. Discard Ratio}
\begin{figure}[!ht]
    \centering
    \includegraphics[width=1\linewidth]{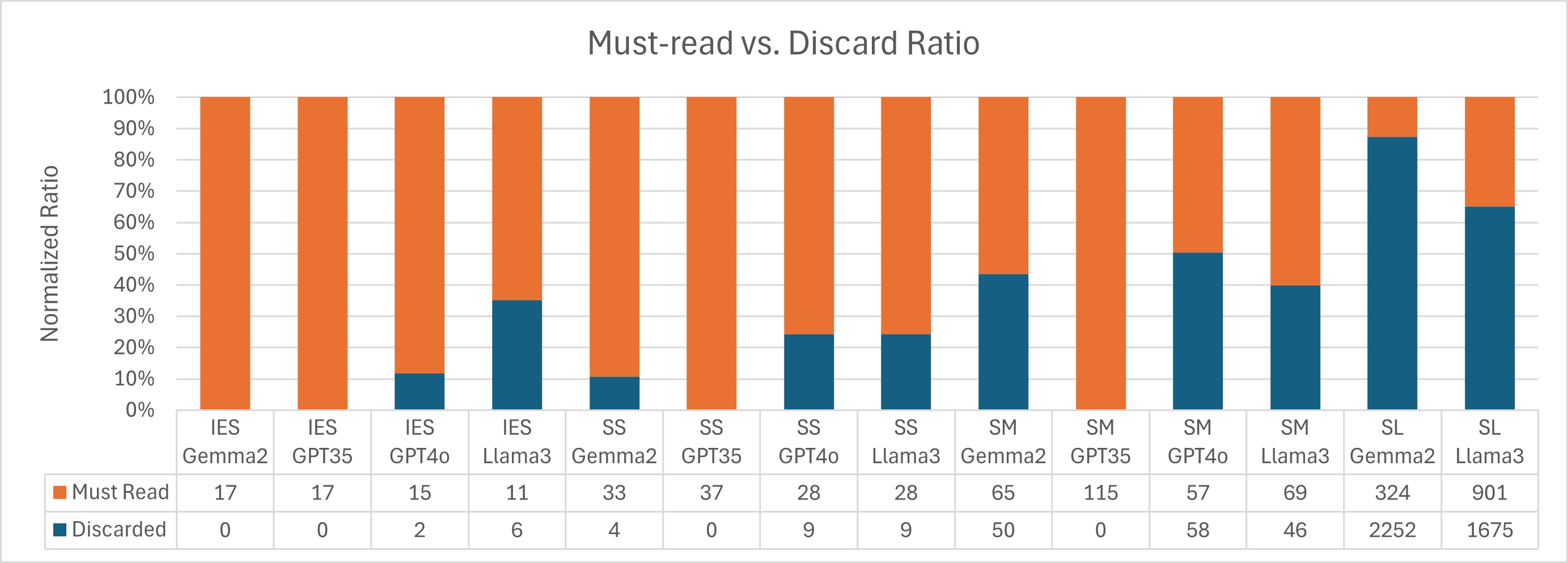}
    \caption{Must-read vs. Discard ratio (all datasets)}
    \label{fig:figure3}
\end{figure}

All small dataset test runs show a relatively low discard ratio, which is expected due to the specificity of the search term, broad research questions, and the low dataset diversity. A comparison between each test run can be seen in Figure \ref{fig:figure3}.

\subsubsection{Reasoning Quality}
A manual review of the system provided coherent explanations for its relevance and contribution assessments, offering insights into why each paper was or wasn't considered relevant to each research question. There is no controlled experiment done to measure quantitatively. The LLMs have to output their reasoning to help the researchers in manually discriminating the articles \citep{Vasconcelos2023Explanations} and can be also part of cognitive forcing functions to be the checkpoint before downstream processing \citep{Buçinca2021To}.

\subsection{Large Dataset Test}
Table 1 shows two types of binary decisions made by LLAssist: the binary relevance and the binary contribution indicator. Referring to the row id: SL-Gemma2 which contains the result of running large dataset test using Gemma 2:9B, the analysis of the larger Scopus dataset (2,576 articles) revealed key insights:

\begin{table}
\centering
\caption{Binary Relevance and Binary Contribution Decisions (all datasets)}
\tiny
\begin{tabular}{|l|c|c|c|c|c|c|c|c|c|c|c|}
\hline
& \multicolumn{1}{c|}{} & \multicolumn{5}{c|}{N Relevant Articles} & \multicolumn{5}{c|}{N Contributing Articles} \\
\cline{3-12}
& N Total & R Any & RQ1 & RQ2 & RQ3 & RQ4 & C Any & RQ1 & RQ2 & RQ3 & RQ4 \\
\hline
IES-Gemma2 & 17 & 17 & 9 & 13 & 1 & 9 & 16 & 8 & 9 & 1 & 3 \\
IES-GPT35 & 17 & 17 & 16 & 17 & 16 & 17 & 17 & 16 & 17 & 16 & 17 \\
IES-GPT4o & 17 & 15 & 11 & 8 & 2 & 2 & 13 & 9 & 6 & 1 & 1 \\
IES-Llama3 & 17 & 10 & 9 & 5 & 1 & 0 & 11 & 10 & 4 & 0 & 2 \\
SS-Gemma2 & 37 & 33 & 25 & 26 & 7 & 12 & 30 & 18 & 19 & 2 & 5 \\
SS-GPT35 & 37 & 37 & 37 & 37 & 33 & 37 & 37 & 37 & 37 & 33 & 37 \\
SS-GPT4o & 37 & 28 & 20 & 16 & 7 & 3 & 24 & 15 & 11 & 5 & 2 \\
SS-Llama3 & 37 & 23 & 19 & 9 & 6 & 1 & 27 & 23 & 12 & 4 & 4 \\
SM-Gemma2 & 115 & 65 & 45 & 50 & 24 & 23 & 46 & 29 & 33 & 5 & 11 \\
SM-GPT35 & 115 & 115 & 111 & 114 & 109 & 114 & 115 & 111 & 114 & 109 & 114 \\
SM-GPT4o & 115 & 57 & 36 & 29 & 23 & 11 & 43 & 24 & 24 & 11 & 6 \\
SM-Llama3 & 115 & 51 & 33 & 19 & 19 & 1 & 63 & 47 & 29 & 12 & 8 \\
\hline
\multicolumn{12}{|l|}{SL-Gemma2*} \\
\hline
Total & 2576 & 324 & 153 & 201 & 110 & 95 & 100 & 27 & 75 & 16 & 26 \\
\hspace{1em}2020 & 301 & 36 & 16 & 21 & 14 & 10 & 4 & 0 & 4 & 1 & 0 \\
\hspace{1em}2021 & 380 & 27 & 11 & 15 & 12 & 7 & 8 & 1 & 6 & 4 & 0 \\
\hspace{1em}2022 & 571 & 52 & 17 & 37 & 15 & 16 & 16 & 0 & 15 & 0 & 7 \\
\hspace{1em}2023 & 869 & 117 & 60 & 69 & 44 & 30 & 42 & 15 & 30 & 6 & 10 \\
\hspace{1em}2024 & 455 & 92 & 49 & 59 & 25 & 32 & 30 & 11 & 20 & 5 & 9 \\
\hline
SL-Llama3 & 2576 & 536 & 387 & 123 & 182 & 7 & 791 & 612 & 231 & 97 & 48 \\
\hline
\multicolumn{12}{l}{* SL-Gemma2 data is broken down by year} \\
\end{tabular}
\end{table}

\begin{enumerate}
    \item \textbf{Trend in Relevance:} There's a notable increase in potentially relevant articles from 2020 to 2023, with a peak in 2023 (869 articles, 117 must-read). The slight decrease in 2024 reflects the mid-year data collection cut-off rather than a decline in research quality. Additionally, identifying the most relevant articles accurately is more important than finding a large number of potentially relevant articles.
    \item \textbf{Research Question Specifics:} RQ2 (risks and vulnerabilities of LLMs in cybersecurity) consistently has the highest number of relevant and contributing articles across years. It indicates it's likely the most well-defined or central question. RQ1 (LLMs for threat detection) shows a sharp increase in relevance from 2022 to 2023. RQ3 (LLMs for adversarial examples/evasive malware) and RQ4 (ethical considerations) have fewer articles but show an upward trend, indicating more specialized areas of increasing importance.
    \item \textbf{Must-Read vs. Contributing Articles:} While 324 articles (12.6\%) are identified as must-read, only 100 articles (3.9\%) are classified as potentially contributing. This suggests that LLAssist is more selective in identifying articles that directly contribute to answering the research questions, implying reduced time to manually read the abstract.
    \item \textbf{Year-over-Year Growth:} The number of potentially relevant articles increased significantly from 2020 to 2023, indicating growing research interest in the field. 2023 stands out as a pivotal year with the highest numbers of potentially high-quality and relevant publications across all categories.
    \item \textbf{Research Question Focus:} RQ2 consistently receives the most attention, suggesting that potential risks and vulnerabilities of LLMs in cybersecurity are a primary concern in the field. RQ4, focusing on ethical considerations, shows the least but growing number of relevant articles, particularly from 2022 onward.
\end{enumerate}

\section{ANALYSIS AND DISCUSSION}
\label{sec:analysis and discussion}
\subsection{Overall Performance}
LLAssist effectively identifies relevant papers, works with various LLM backends, and significantly reduces manual screening time. On the other hand, the system did not utilize all available metadata (e.g., publication year, citation counts) in its relevance assessments, which could have provided additional context. Also, different LLMs behave differently, necessitating more precise prompt tuning. The analysis was also limited to titles and abstracts, potentially missing relevant information contained in the full text of the papers.

\subsection{Time and Cost Efficiency}
LLAssist's throughput varies across models and dataset sizes. It processes datasets of 17-37 articles in under 10 minutes, 115 articles in 20-50 minutes, and 2,576 articles in 10-11 hours. Among the models tested, GPT4o emerges as the slowest, processing articles in 24-29 seconds on average. Llama3 is the fastest, consistently quick at 10-11 seconds per article. Gemma2 and GPT35 offer similar speeds, averaging 12-14 seconds for each article processed. The per-article processing times remain consistent across dataset sizes, indicating scalability. This is a significant improvement over human performance \citep{Wallace2010Semi-automated,Joos2024Cutting}.

Cost-wise, GPT-4o is the most expensive at approximately \$3.16 per 100 articles while GPT-3.5 offers a more budget-friendly option at about \$0.22 per 100 articles. Meanwhile, both Gemma 2 and Llama 3 do not have a set cost due to the ability to run locally without cloud services. Notably, the high discrimination ability of Gemma 2 may help researchers to do the initial screening of many articles without relying on cloud services.

\section{FUTURE WORK}
\label{sec:future work}
LLAssist's limitations include dependence on LLM quality and input formatting, focus on titles and abstracts, and potential misalignment with human judgment. Future work should aims to incorporate full-text analysis, implement feedback mechanisms, and develop domain-specific models for improved accuracy.

\section{CONCLUSION}
\label{sec:conclusion}
In conclusion, LLAssist demonstrated promising capabilities in automating the initial stages of a literature review. Its ability to quickly process and categorize papers offers valuable support to researchers. However, there is room for improvement in utilizing more of the available metadata and fine-tuning the relevance criteria to better differentiate between highly relevant and marginally relevant papers. By providing a simple, transparent tool for literature review automation, LLAssist not only enhances research efficiency but also promotes open science principles and methodological freedom in the rapidly evolving landscape of AI-assisted research. While not a replacement for human judgment, LLAssist can significantly reduce the time spent on initial screening and help researchers focus their efforts on the most promising and relevant articles for their research questions, allowing researchers to focus on high-quality work, to achieve higher productivity across expertise and industry.

\section{AVAILABILITY}
\label{sec:availability}
The source code for LLAssist is freely available at \url{https://github.com/cyharyanto/llassist} and is being actively developed to improve usability. We encourage researchers to use, modify, and contribute to this tool to further advance the efficiency of academic literature reviews across various disciplines.

\section{ACKNOWLEDGMENT}
\label{sec:acknowledgment}
We would like to express our deep gratitude to Dr. Arathi Arakala, Dr. Argho Bandyopadhyay, and Dr. Jessica Helmi from RMIT University for their invaluable guidance in systematic literature review methodologies.

\newpage

\bibliographystyle{plainnat}
\bibliography{cie-refs}

\end{document}